\documentclass[12pt]{iopart}

\usepackage[numbers,sort&compress,square]{natbib}
\usepackage{graphicx}
\usepackage{subcaption}
\begin{document}

\title[Microseismic Noise Mitigation with Machine Learning for Advanced LIGO]{Microseismic Noise Mitigation with Machine Learning for Advanced LIGO}

\author{Christina Reissel$^{1,2,7}$, Devin Lai$^3$, Shivanshu Dwivedi$^4$, Edgard Bonilla$^3$, Claudia Geer$^4$, Christopher Wipf$^5$, Richard Mittleman$^1$, Philip Harris$^{2,7}$, Eyal Schwartz$^4$, Dovi Poznanski$^{3,6}$, Brian Lantz$^3$, Erik Katsavounidis$^{1,2,7}$}

\address{$^1$ LIGO Laboratory, Massachusetts Institute of Technology, Cambridge, MA 02139, USA \\ 
$^2$ Department of Physics, MIT, Cambridge, MA 02139, USA \\
$^3$ Department of Physics, Stanford University, 382 Via Pueblo Mall, Stanford, CA 94305, USA \\  
$^4$ Trinity College, Physics Department, Hartford, CT, 06106, USA \\ 
$^5$ LIGO Laboratory, California Institute of Technology, Pasadena, CA, USA\\
$^6$ School of Physics and Astronomy, Tel Aviv University, Tel Aviv 69978, Israel\\
$^7$ NSF AI Institute for Artificial Intelligence and Fundamental Interactions (IAIFI)}



\begin{abstract}
The unprecedented sensitivity of the Laser Interferometer Gravitational-Wave Observatory, which enables the detection of distant astrophysical sources, also renders the detectors highly susceptible to low-frequency ground motion. Persistent microseisms in the 0.1--0.3~Hz band couple into the instruments, degrade lock stability, and contribute substantially to detector downtime during observing runs. The multi-stage seismic isolation system has achieved remarkable success in mitigating such disturbances through active feedback control, yet residual platform motion remains a key factor limiting low-frequency sensitivity and duty cycle. Further reduction of this residual motion is therefore critical for improving the long-term stability and overall astrophysical reach of the observatories.

In this work, we develop a data-driven approach that uses machine learning to model and suppress residual seismic motion within the isolation system. Ground and platform sensor data from the detectors are used to train a neural network that predicts platform motion driven by microseismic activity. When incorporated into the control scheme, the network’s predictions yield up to an order-of-magnitude reduction in residual motion compared to conventional linear filtering methods, revealing that non-linear couplings play a significant role in limiting current isolation performance. These results demonstrate that machine-learning-based control can provide a powerful new pathway for enhancing active seismic isolation, improving lock robustness, and extending the low-frequency observational capabilities of gravitational-wave detectors.
\end{abstract}

%
%
%
%
%

\section{Introduction}
\label{sec:introduction}

With the first direct detection of gravitational waves in 2015~\cite{LIGOScientific:2016aoc}, 
the Laser Interferometer Gravitational-Wave Observatory (LIGO) not only confirmed one of the central predictions of Einstein’s theory of general relativity but also inaugurated a new era of gravitational-wave astronomy. 
The ability of detectors like LIGO to make these discoveries depends critically on their extreme sensitivity to minute distortions, on the order of $10^{-19}$\,m.
However, this extraordinary sensitivity also makes LIGO highly susceptible to environmental disturbances, particularly at frequencies below 20\,Hz  where seismic motion, and in particular the noise of the interferometer control loops to compensate for that motion, represent the leading contribution to the detector noise budget~\cite{Capote:2024rmo}. 

Among these disturbances, microseisms, continuous, low-frequency vibrations typically between 0.05 and 0.3\,Hz, pose a great challenge to the detector stability and duty cycle.
Generated mainly by interactions between ocean waves and the seafloor or coastline, microseisms can also be influenced by distant earthquakes and local anthropogenic activity. 
While their frequencies are far below the main detection band, microseisms impact the detector both directly and indirectly: they can couple into higher-frequency bands through non-linear mechanisms~\cite{Soni:2023kqq}, degrade interferometer alignment and limit lock stability. 
For the O4 observing run, our analysis finds that microseismic and teleseismic ground motion contributed approximately $8$--$9\%$ of the total detector downtime. Similar analyses for O2 and O3 yield $5$--$9\%$. Taken together, these results indicate a persistent and gradually increasing difficulty in maintaining detector stability during intervals of elevated ground motion, underscoring their importance for overall observatory performance. Any reduction in seismic noise coupling in this regime could therefore yield broad improvements in duty cycle, calibration precision, and sensitivity to astrophysical events.

To mitigate these effects, advanced LIGO employs a multi-stage seismic isolation and suspension system designed to decouple the interferometer optics from ground motion~\cite{Matichard:2015eva}. 
The \textit{Seismic Isolation System} (SEI) consists of external hydraulic and in-vacuum active isolation stages, 
which together monitor and counteract ground vibrations across a wide frequency range. 
The \textit{Suspension System}, comprising multi-stage pendulums that support the core optics, provides additional passive filtering at higher frequencies. 
The combined SEI and suspension hierarchy achieves several orders of magnitude of attenuation between the ground and the interferometer optics, 
reducing ground motion by more than ten orders of magnitude at 30\,Hz
and enabling sub-nanometer stability.
This work focuses on the active seismic isolation system. 
The SEI relies on multiple inertial and position sensors whose outputs are linearly combined to estimate the platform motion along each degree of freedom. 
Through a network of \textit{active feedback control loops}, corrective forces are applied to counteract detected disturbances in real time~\cite{Schwartz20}.
\begin{figure}[h]
    \centering
    \includegraphics[width=0.8\linewidth]{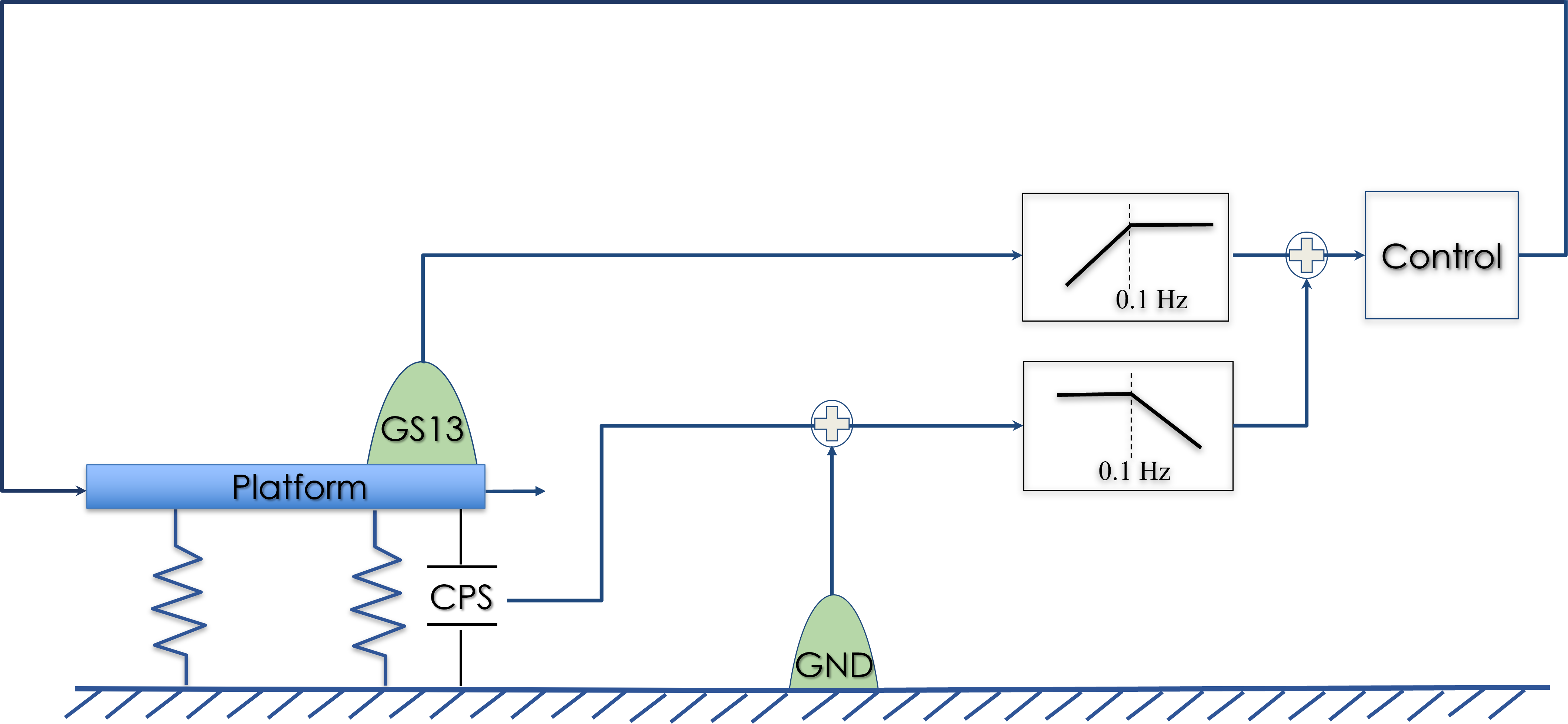}
    \caption{Illustration of the HAM seismic controls. Sensors on the ground and on the platform are combined to act as a witness of the platform’s inertial motion. Control signals are derived and sent to actuators on the platform.}
    \label{fig:1}
\end{figure}
In Figure \ref{fig:1}, we illustrate a typical seismic controls of the Horizontal Access Module (HAM) which house the interferometer’s auxiliary optics. The figure also illustrates how the multitudes of sensors are combined to create an absolute inertial measurement of the platform motion. 

Despite its success, the performance of the isolation system remains limited by \textit{residual cross-couplings} 
between translational and rotational degrees of freedom. 
A long-standing hypothesis attributes these to \textit{tilt-horizontal coupling}, 
in which sensor miscalibration or platform tilt converts vertical motion into apparent horizontal displacement~\cite{Lantz_rotation}. 
Several calibration and tuning campaigns have aimed to reduce these couplings in LIGO~\cite{Carter:2020kuf, Tsang:2021cgz, GS13alog} and other gravitational-wave observatories~\cite{Tsang:2024rke,Saffarieh:2025ahl}, 
for example by adjusting sensor-correction filters, refining cross-coupling matrices, and optimizing sensor blends. 
However, these efforts have yielded only modest improvements, suggesting that the remaining coupling mechanisms are more complex 
and possibly non-linear~\cite{nonlinear}, potentially arising from sensor saturation, structural asymmetries, or non-stationary environmental conditions.

In this work, we explore an alternative, data-driven approach based on \textit{machine learning} (ML) techniques. Previous ML approaches focus on predicting the seismic activity directly from ground motion, but explored only non-causal models~\cite{Basalaev:2024vsb}. Our work focuses on the development of a causal ML model whose output signal can be fed into the existing seismic control system.  
We model the platform motion using the various sensors  
in the SEI, extending beyond a linearized model without relying on simplified physical assumptions. 
We demonstrate that these methods can substantially reduce residual platform motion induced by microseismic activity. Recent advances in angular control at LIGO Livingston demonstrate the substantial performance gains that machine learning can achieve when integrated directly into the control system~\cite{TheLIGOInstrumentTeam:2025xha}. 

This paper is structured as follows: Section~\ref{sec:dataset} describes the setup and different sensor outputs used as inputs in our studies. Section~\ref{sec:methods} describes the various methods we explored to learn the residual motion from said sensor outputs, including linear and non-linear ML methods. Section~\ref{sec:implementation} describes how these approaches can be integrated into the LIGO control system. In Section~\ref{sec:summary}, we summarize our findings and describe potential future work and improvements. 

\section{Dataset}
\label{sec:dataset}

The LIGO \textit{Seismic Isolation sub-system} (SEI) consists of two types of isolation platforms: one designed for the larger vacuum chambers that house the test mass mirrors and beamsplitters, and another for the smaller ‘HAM’ vacuum chambers, which contain additional optics such as recycling mirrors, mode-cleaner mirrors, and mode-matching telescopes. We use data from the HAM5 table installed in the LIGO-Livingston detector taken during the O4a run. 

Several sensors measure the current motion of the table along the different degrees of freedom: Geotech GS-13 seismometers mounted on the table (referred to as ‘GS13’) sense vibrations, while capacitive displacement sensors (CPS) measure relative motion between the table and its support structure. An additional seismometer (typically a Streckeisen STS-2 or equivalent, labelled ‘GND’) is placed on the floor to sense ground motion.
The ground (GND) sensors measure motion in the three translational degrees of freedom (DOF) (X, Y, Z). The capacitive displacement sensors and the GS13s measure of all DOFs of the table: the vertical and horizontal position (X, Y, Z) as well as the rotation around those axes (RX, RY, RZ). A schematic overview of the different sensors and their placements can be found in Figure~\ref{fig:1}.

For ideal sensors (perfect calibration, noise-free, and free from cross-couplings), 
we expect the measurements in each direction to agree, i.e., $GS13_i=CPS_i-GND_i$. 
Because the sensors are used to control the platforms, errors in the sensors will degrade the isolation performance.
Optimizing the isolation therefore corresponds to minimizing the residual between the left and right hand sight of these equations. For convenience, we cast the problem as one of regressing the $GS13_i$ vectors. The output of the CPS and GND sensors accounts for nine time series in total. We use these nine time series to predict the resulting table motion, as it is monitored by the GS13 sensor. Ultimately, the predicted (residual) motion would be fed to the actuators to cancel the table motion, leading to a better active microseismic noise suppression.

\section{Methods}
\label{sec:methods}
\subsection{Linear Analysis}
\label{sec:linear}
Using the aforementioned HAM datasets, we fit linear models to provide a baseline performance measurement. We apply three different filters: Non-causal top-hat filter, non-causal filter with roll-off of $f^2$, and a causal filter. We use linear regression to predict the six GS13 degrees of freedom with the CPS, GND, and non-target GS13 directions as inputs.

\begin{figure}[h!]
  \centering
  \includegraphics[width=\textwidth]{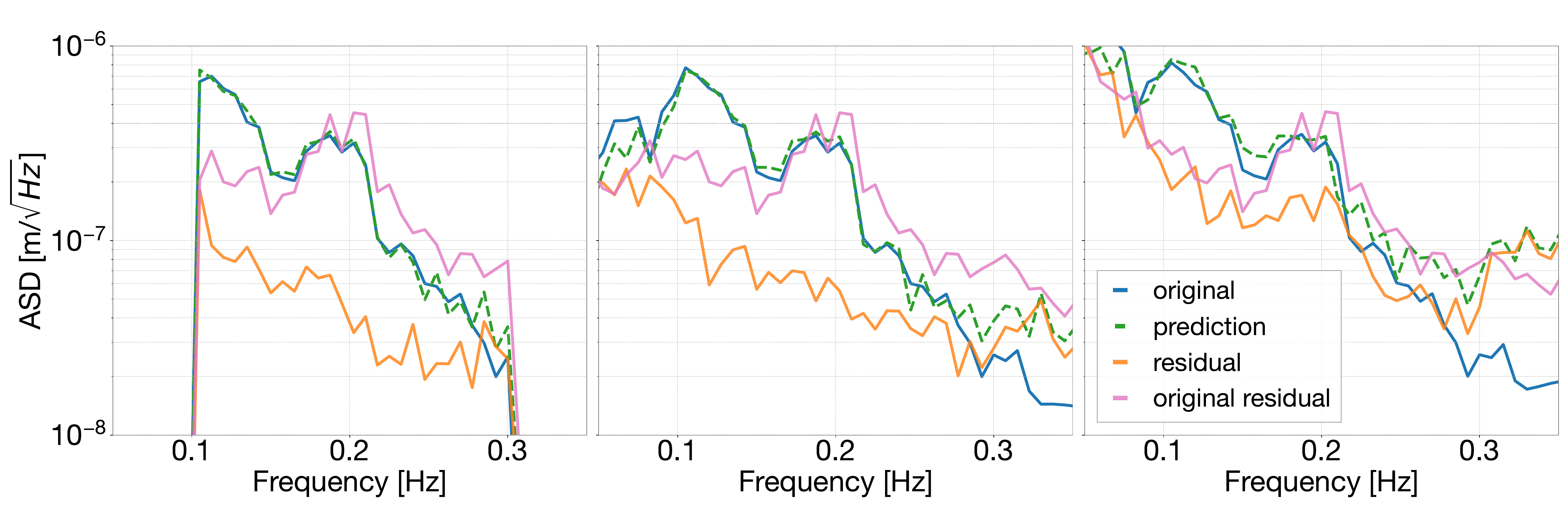}
  \caption{Amplitude spectral density (ASD) of the original residual table motion GS13 in X-direction (blue) compared to the prediction based on the sensor outputs from the linear model (green) and residual between the motion and the prediction (orange). All data is taken on October 16, 2023 (GPS time 1381528818) and calibrated with a top-hat function (a), f2 filter (b) and causal filter (c).}
  \label{fig:performance}
\end{figure}

Figure \ref{fig:performance} shows the fit of the GS13 signal in X-direction when the data are bandpassed with the three different filters. We find the linear regression to be able to predict the target GS13 signals, particularly when bandpassed via the top-hat filter which achieves residuals are nearly an order of magnitude lower than the original residuals. However, as expected, the causal filter performs worse than the other filters in the band of interest. Regardless of the filtering method, any improvement in the frequency band of interest comes at the expense of increased noise at higher frequencies, which will require further filtering before feeding any correction to the actuators. 

\begin{figure}[h!]
  \centering
\includegraphics[width=\textwidth]{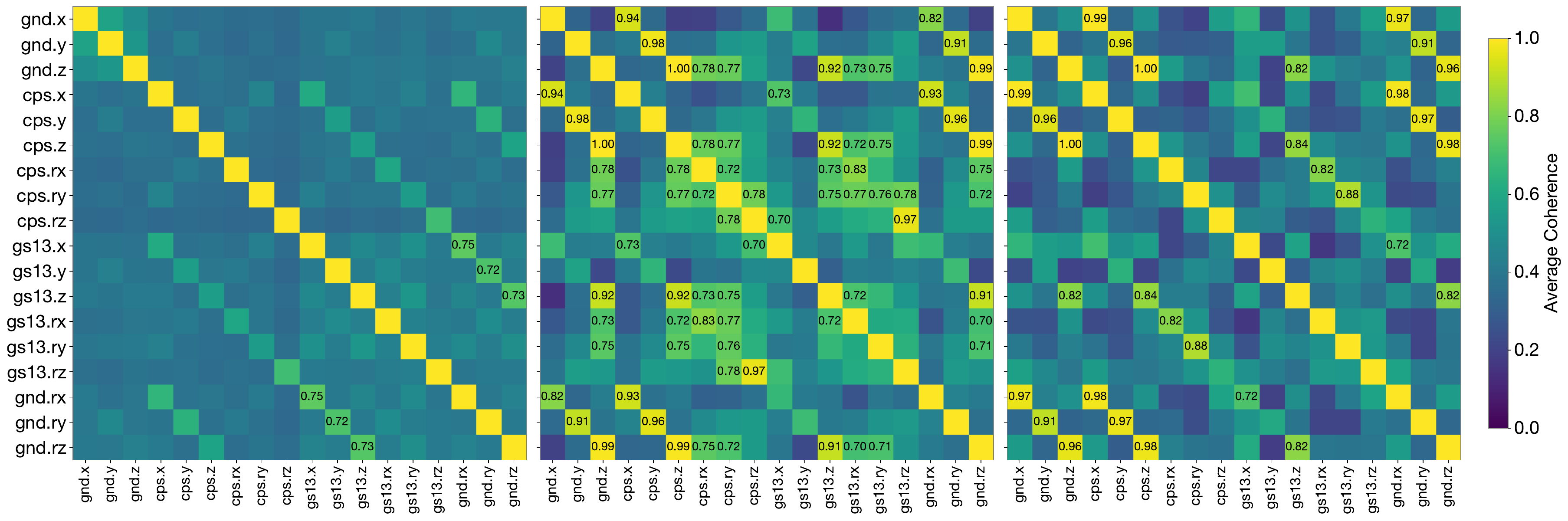}
  
\caption{Average coherence between the channels, between 0.1 and 100Hz  (a), vs. 0.1--0.3Hz at one time (b), and a different time (c). The system appears to show time dependent cross couplings in the microseismic band.} 
  \label{fig:cov}
\end{figure}

\begin{figure}[h!]
\centering
  \includegraphics[width=0.8\textwidth]{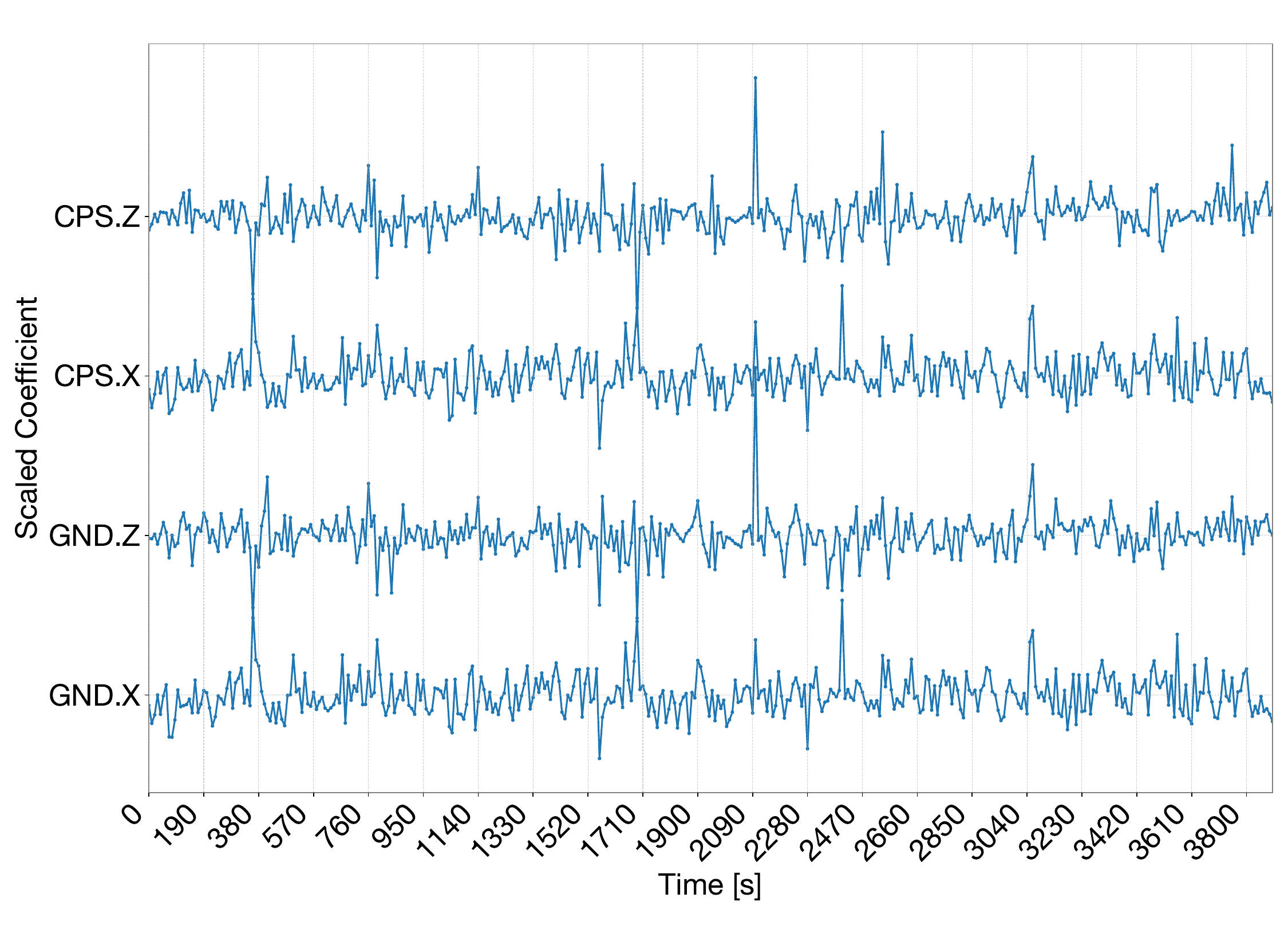}
  \caption{Time dependence of a some significant coefficients of channels as predictors of GS13X.}
  \label{fig:spike}
\end{figure}

Additionally, we find the parameters of the fit to be unstable, time dependent, and frequency dependent. This can be see, for example, looking at the coherence matrices between the channels during large microseismic activity, which we show in Figure~\ref{fig:cov}. While a broad frequency window (0.1--100Hz, left) reveals a relatively simple picture that fits our expectations of the dependence between the channels, a narrower frequency window (0.1--1Hz, middle) reveals cross coupling between the different degrees of freedom. Worse, the same narrow window, but at a different time (right panel), reveals significantly different correlations. This can be further seen in Figure~\ref{fig:spike}, where we perform a new linear fit every 10 seconds, and plot the value for some of the best fit parameters, showing they all fluctuate wildly. These results demonstrate clear limitations in the use of a linear subtraction strategy as a means to mitigate seismic effects. 

\subsection{Machine Learning}
\label{sec:training}
While we find a linear model to improve the residuals, our results in Section~\ref{sec:linear} indicate that a full solution requires a more flexible framework. To cancel the remaining residual motion, we train a Long Short-Term Memory (LSTM) network to predict the next time step of the table motion, using only the last 60\,s of the table motion's monitoring GS13 sensor as inputs. LSTMs are recurrent neural networks designed to learn long-term dependencies in sequential data by using special memory cells and gating mechanisms. Our choice for LSTMs is motivated by the fact that they are well understood and simple to implement and have been successfully been used for similar tasks. While alternative approaches like Convolutional Neural Networks (CNNs) are frequently deployed in  

All results shown in this section focus on predicting the table motion in X-direction recorded with the GS13 sensor. For optimal performance, we apply additional preprocessing to the sensor data described in Section~\ref{sec:dataset}: We resample all outputs to 4\,Hz and apply a Butterworth bandpass filter (which has a causal infinite impulse response) to 0.1 to 0.3\,Hz to emphasize the frequency band in which microseismic activity dominates and reduce the impact from higher frequency noise.
The main dataset is taken on October 16, 2023 and consists of 3000\,s that are split sequentially into a training, validation and test dataset (0.6:0.2:0.2). 

The model consists of three LSTM layers with a hidden size of 128. The trainable weights in the model are optimized by applying the mean squared error (squared L2 norm) criterion between each element in the time series of the monitored and predicted residual motion. 

To feed the consecutive time series into the neural network, we divide the data into overlapping segments where each segment is 60\,s long with an offset of 0.25\,s (one time step) between segments. The batch size, e.g., the number of training samples used for a single forward and backward pass, is 32. Parameters are updated using the \texttt{Adam} optimizer. The learning rate, the hyperparameter that controls how much the model's weights are adjusted during training in response to the calculated error, is set to $10^{-3}$. 



\begin{figure}[h!]
    \centering
    \begin{subfigure}[t]{0.49\textwidth}
        \centering
        \includegraphics[width=\textwidth]{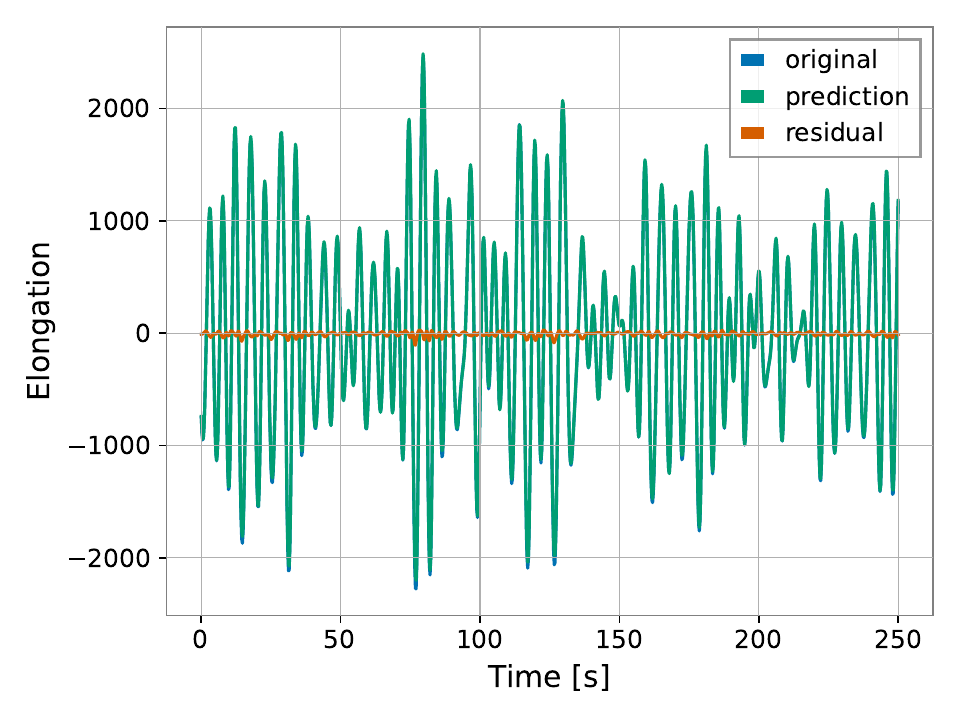}
        \caption{GS13 sensor output (X-direction).}
        \label{fig:performance_a}
    \end{subfigure}
    \hfill
    \begin{subfigure}[t]{0.49\textwidth}
        \centering
        \includegraphics[width=\textwidth]{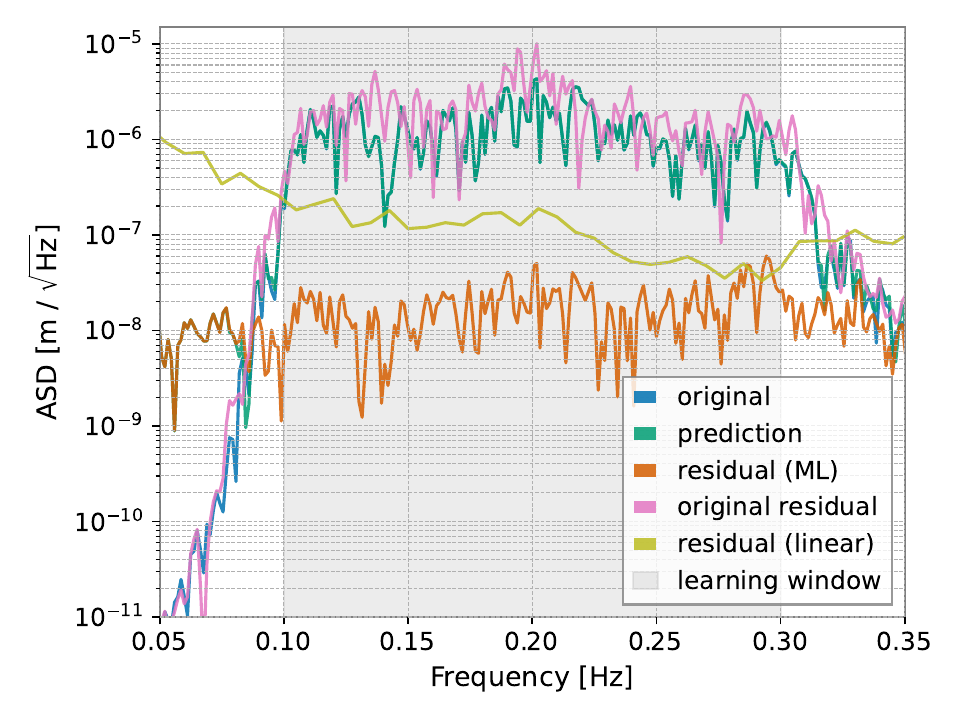}
        \caption{Amplitude spectral density (ASD) for GS13 output (X-direction).}
        \label{fig:performance_b}
    \end{subfigure}
    
    \caption{GS13 sensor output in horizontal (X) direction (a) and corresponding Amplitude spectral density (ASD). The original measured residual table motion (blue) is compared to the predicted table motion (green). The orange line shows the difference between the actual motion and the prediction, indicating the possible improvement if the neural network output is used for active feedback control over the current residual (pink). The grey area indicates the frequency band used during training.}
    \label{fig:two-subfigs}
\end{figure}

All results are obtained using the independent test dataset, which is not utilized during the training. Figure~\ref{fig:performance_a} shows the residual motion as predicted with the neural network compared to the actual residual motion in the horizontal X direction monitored by the GS13 sensor for 1000\,s of the independent test dataset that was not utilized during the training. We find that the prediction closely resembles the actual measured motion. Figure~\ref{fig:performance_b} shows the amplitude spectral density (ASD) for the predicted and measured residual table motion with special emphasis on the frequency band between 0.1 and 0.3\,Hz in which the microseismic noise is dominant and which corresponds to the dataset we utilized during training. The neural network prediction also closely resembles the actual table motion in the frequency domain. The ratio between the measured and predicted table motion indicates the potential improvement if the predicted motion is subtracted from the actual table motion. The ratio suggests improvements two order of magnitudes in frequency bands in which microseism is most prevalent. 

\begin{figure}[h!]
  \centering
\includegraphics[width=0.8\textwidth]{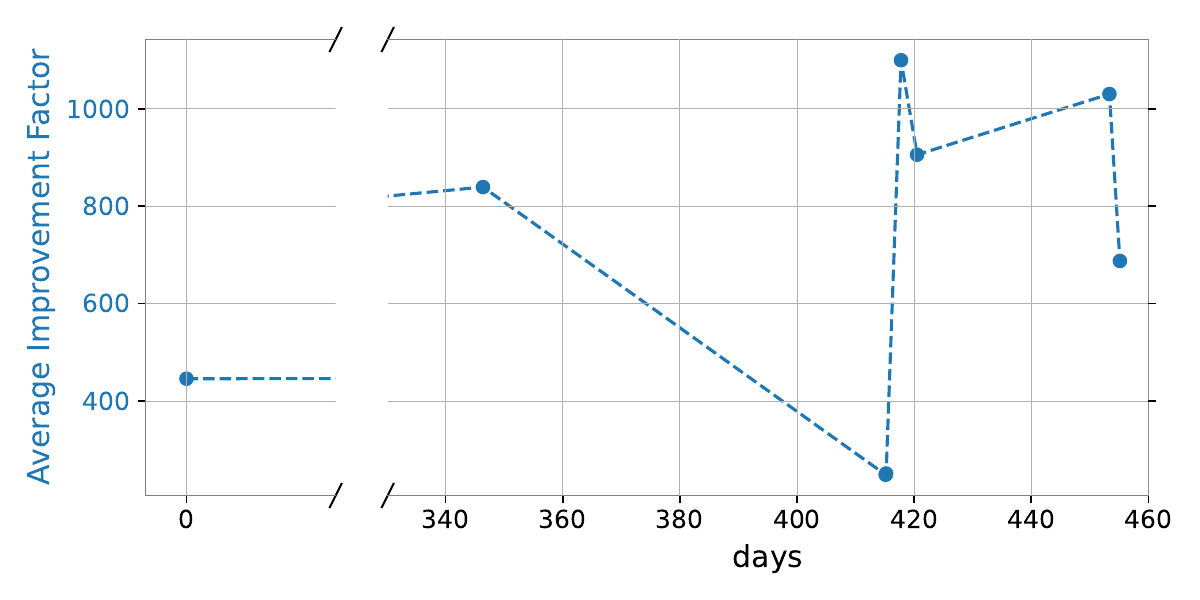}
\caption{Average Improvement Factor measured as ASD ratio between the GS13 sensor output (X-direction) and the residual obtained after subtracting the neural network output. The time difference is the time between training and evaluation datasets. We consider only the relevant frequency band between 0.1 and 0.3\,Hz for the average.}
  \label{fig:performance_late}
\end{figure}

In addition to the results from the machine learning algorithm, Figure~\ref{fig:performance} contains also the results obtained by the linear methods as described in Section~\ref{sec:linear}. 
We find the neural network approach outperforms the linear model by up to a factor of ten, indicating that an essential part of the residual motion is caused by non-linear cross couplings, which the neural network, as a non-linear regression method, is able to capture.

To test how well the neural network generalizes, we evaluate the network on data taken more than one year after the training dataset. We measure the average improvement factor, e.g. the ASD ratio of the GS13 sensor output to the new residual obtained from subtracting the neural network prediction in the frequency band between 0.1 and 0.3\,Hz, in dependence of the time shift to the training dataset. The result can be found in Figure~\ref{fig:performance_late}. 
While we find some fluctuations in the performance which are most likely related to changes in the environmental conditions, we find the improvement factor consistently to be larger than 100. 


\section{Implementation}
\label{sec:implementation}
There are several ways that our findings can be used to improve the isolation performance, including direct recalibration of individual sensors, linear corrections of the sensor outputs, and non-linear correction of sensors. 

Correcting the sensors' calibration would be the most straightforward implementation. For example, the rotation signals are generated as the difference between various linear sensors - and so if the calibration of the sensors is slightly different, then the relatively large translation signals can generate a spurious rotation response. If the linear models can identify persistent residuals from mis-calibrations, these can be directly corrected in the calibration blocks of the realtime controls. As mentioned in the Section~\ref{sec:introduction}, ongoing work of this type using simpler methods has not been successful. 

Linear corrections of the residual is a second method to implement these corrections. Once a linear regression has been found which predicts the residual for a GS13 channel from the other sensor channels, we can implement that regression with the LIGO control computers. The residual signal can be generated in realtime with existing filter tools and matrix combinations. The residual can then be  subtracted from the relevant GS13 signal, and the corrected signal used for platform control. A prototype version of this is now running on the isolation prototype at Stanford. This implementation is challenged by three issues - it can only correct linear couplings between sensors; it assumes that the cross-couplings are stable with time; and although the cross-coupled signals of the sensors can be subtracted, the sensors' noise will add incoherently and may limit the implementation. 

Similar to applying linear corrections, machine learning models can also be used to apply non-linear corrections. To generate these non-linear corrections in real time, we extract the trained neural network into \texttt{C} code and deploy it within the LIGO control system. Our initial tests with networks similar to the one used in this work demonstrate that the extraction and integration are feasible, fulfilling the timing constraints of the system. Future work on the prototypes will focus on actual testing this non-linear feedback in hardware, particularly under a variety of settings to ensure stability to changes in the environmental conditions.

\section{Summary and Outlook}
\label{sec:summary}
We investigate residual seismic noise in LIGO’s active SEI system, focusing on motion in the microseismic band. As a baseline, conventional linear filtering techniques are applied, yielding measurable improvements in residual platform motion. However, the persistence of strong cross-couplings, particularly between translational and rotational degrees of freedom, demonstrates that linear models alone cannot fully capture the system dynamics.

To address these limitations, we implement an LSTM network trained to predict residual table motion from the platform sensor data. The network significantly outperforms linear models, achieving up to two orders of magnitude reduction in the ASD of residual motion within the dominant microseismic band. The model also exhibits partial generalization to data acquired more than a year after training, suggesting the feasibility of real-time deployment.

These results confirm that non-linear couplings play a substantial role in limiting the current performance of the SEI. Incorporating machine-learning-based prediction into the real-time control infrastructure could improve lock stability, reduce downtime during periods of high microseismic activity, and enhance duty cycle and low-frequency sensitivity. To the best of our knowledge, this is the first work to present a causal ML model capable of predicting seismic motion in gravitational-wave detectors in a manner suitable for real-time active feedback control.

Future work will focus on real-time implementation within LIGO prototypes to evaluate closed-loop stability and latency constraints. Adaptive or continuously retrained models will also be explored to account for long-term non-stationarities such as sensor drift, seasonal variations, and hardware modifications. Although this study focuses on the microseismic band, extending the method to other frequency regimes holds great promise to further improve the seismic isolation performance.

Ultimately, combining physical insight with data-driven modelling can enable a new generation of hybrid seismic control strategies, providing significant performance gains for both current and next-generation gravitational-wave observatories.

\section{Acknowledgements}
\label{sec:Acknowledgements}
This material is based upon work supported by NSF's LIGO Laboratory which is a major facility fully funded by the National Science Foundation. This work is further supported by NSF grant NSF-2309161. C.R. acknowledges support under a Swiss National Science Foundation (SNSF) Postdoc.Mobility fellowship (grant number 222340). R.M. and E.K. acknowledge support from the NSF under award PHY-1764464 and PHY-2309200 to the LIGO Laboratory. E.K also acknowledges NSF support under award PHY-2117997 to the A3D3 Institute.
This research was funded in part by the Koret Foundation, the Kavli Institute for Particle Astrophysics and Cosmology at Stanford University, and by grant NSF PHY-2309135 to the Kavli Institute for Theoretical Physics (KITP). D.P. acknowledge support from Israel Science Foundation (ISF) grant 541/17, and by grant 2018017 from the United States-Israel Binational Science Foundation (BSF). Part of the computations in this paper were run on the
FASRC Cannon cluster supported by the FAS Division of Science Research Computing Group at
Harvard University.
This document has been assigned LIGO document number LIGO-P2500700.

\bibliographystyle{my-iopart-num}
\bibliography{sample.bib}

\end{document}